\documentclass[namedreferences]{SolarPhysics}
\usepackage[optionalrh,solaenum]{spr-sola-addons}
\usepackage{graphicx}                  
\usepackage{amssymb}                    
\usepackage{color}                       
\usepackage{url} 
 
\begin{document}
\begin{article}
\begin{opening}
\title{Visibility graph analysis of solar wind velocity}
\author{Vinita~\surname{Suyal}$^1$\sep
        Awadhesh~\surname{Prasad}$^2$\sep
        Harinder~P.~\surname{Singh}$^3$
       }
\runningauthor{Suyal et al.}
\runningtitle{Visibility graph analysis of solar wind velocity}
\institute{$^{1,2,3}$Department  of Physics  and Astrophysics, University of Delhi, Delhi 110007, India.\\
email:$^1$vinita.suyal@gmail.com\\
$^2$awadhesh@physics.du.ac.in\\
$^3$hpsingh@physics.du.ac.in}
\begin{abstract}
We analyze \textit{in situ} measurements
of solar wind velocity obtained by
Advanced Composition Explorer (ACE) spacecraft and
Helios spacecraft during the years $1998-2012$ and
$1975-1983$ respectively. The data belong to mainly solar
cycle $23$ ($1996-2008$) and solar cycle $21$
($1976-1986$) respectively. We use Directed Horizontal Visibility
graph (DHVg) algorithm and estimate a graph
functional, namely, the degree distance (D) as the Kullback-Leibler
divergence (KLD) argument to understand time irreversibility
of solar wind time series.
We estimate this degree distance irreversibility
parameter for these time series
at different phases of solar activity
cycle. Irreversibility parameter is first established
for  known dynamical data and then applied for solar wind velocity time series.
It is observed that irreversibility
in solar wind velocity fluctuations
show similar behavior at $0.3$ AU (Helios data) and $1$ AU (ACE data).
Moreover it changes over the different phases of  solar activity cycle.
\end{abstract}
\end{opening}
\section{Introduction}
\label{S-intro}
Solar wind is the continuation of the solar corona
and it continuously expands into space with speeds
that can vary from about $250$ km/s to more than
$800$ km/s \cite{schw07}.
The interaction between solar wind and the local interstellar
medium determines the size and boundaries of the heliosphere \cite{balo08}.
Solar wind is a turbulent plasma and this
turbulence is anisotropic with respect the mean magnetic field
\cite{robi71, sheb83, gold95}.
Solar wind variability (velocity, proton density,
temperature, and helium content) leads
to evolving, dynamic phenomena throughout the heliosphere
on all temporal and spatial scales \cite{gold95a}.
Solar wind velocity shows a high correlation
with coronal holes area \cite{rott12}.
These coronal holes are long-lived structures on the Sun, which
may persist for several solar rotations.

Solar wind has a three-dimensional structure
and it is highly dependent upon solar cycle \cite{hapg91,mcco03,schw07,balo08}.
Its characteristics vary over a solar cycle,
but it maintains its simplest configuration
during solar activity minimum,
when large polar coronal holes dominate the outer solar
atmosphere away from the equator \cite{hans12}.
Due to overlapping of fast streams from the coronal holes
at the solar maximum, solar wind velocity time series has larger fluctuations
\cite{kats12, suya12_a}.

The analysis of solar wind plasma has been an
area of considerable research interest in recent past.
Various complexity measures such as entropy
(\opencite{mace97}, \citeyear{maceo98};
\opencite{mace00}; \opencite{reda01}; \opencite{suya12_a}),
Lyapunov exponents (\opencite{maceo98}; \opencite{reda01}; \opencite{gupt08})
and correlation dimension (\opencite{mace97}; \opencite{gupt08})
show that solar wind velocity fluctuations are a consequence of complex
nonlinear dynamical processes.
\opencite{mila04} suggested an anisotropy in the velocity,
magnetic, and cross helicity
correlation function and power spectra by analyzing magnetic
and bulk velocities measured by ACE.
\opencite{gupt08} analyzed the solar wind velocity data
to observe the inherent changes in the dynamics governing
the solar wind at $0.3$ AU.
\opencite{suya12_b} analyzed the correlation of solar wind velocity
at different phases of the solar activity cycle. They suggested
hysteresis in the dynamics governing the
solar wind over a complete solar activity cycle \cite{suya12_a}.

In the present work, we analyze
the solar wind velocity time series at
different phases of solar activity cycle.
We use the data obtained from ACE during $1998-2012$
and Helios spacecraft during $1975-1983$.
ACE data belong mainly to the solar activity cycle $23$,
while data obtained from Helios spacecraft correspond
to solar activity cycles $21$.
We estimate the time irreversibility
parameter, $D$ using directed horizontal visibility graph
(DHVg) algorithm \cite{laca12} for the solar wind
time series. In the next section, we review the DHVg algorithm
to calculate $D$ of time series.
In Section~\ref{S-rev_irrev} we estimate $D$ for known
dynamical systems. In Section~\ref{S-solar_wind}, we analyze
solar wind data using DHVg technique.
It is followed by  conclusions in Section~\ref{S-conclusion}.
\section{Directed horizontal visibility graph algorithm}
\label{S-visibility_graph}
\begin{figure}
\centering
\vspace{2cm}
\includegraphics[scale=.5]{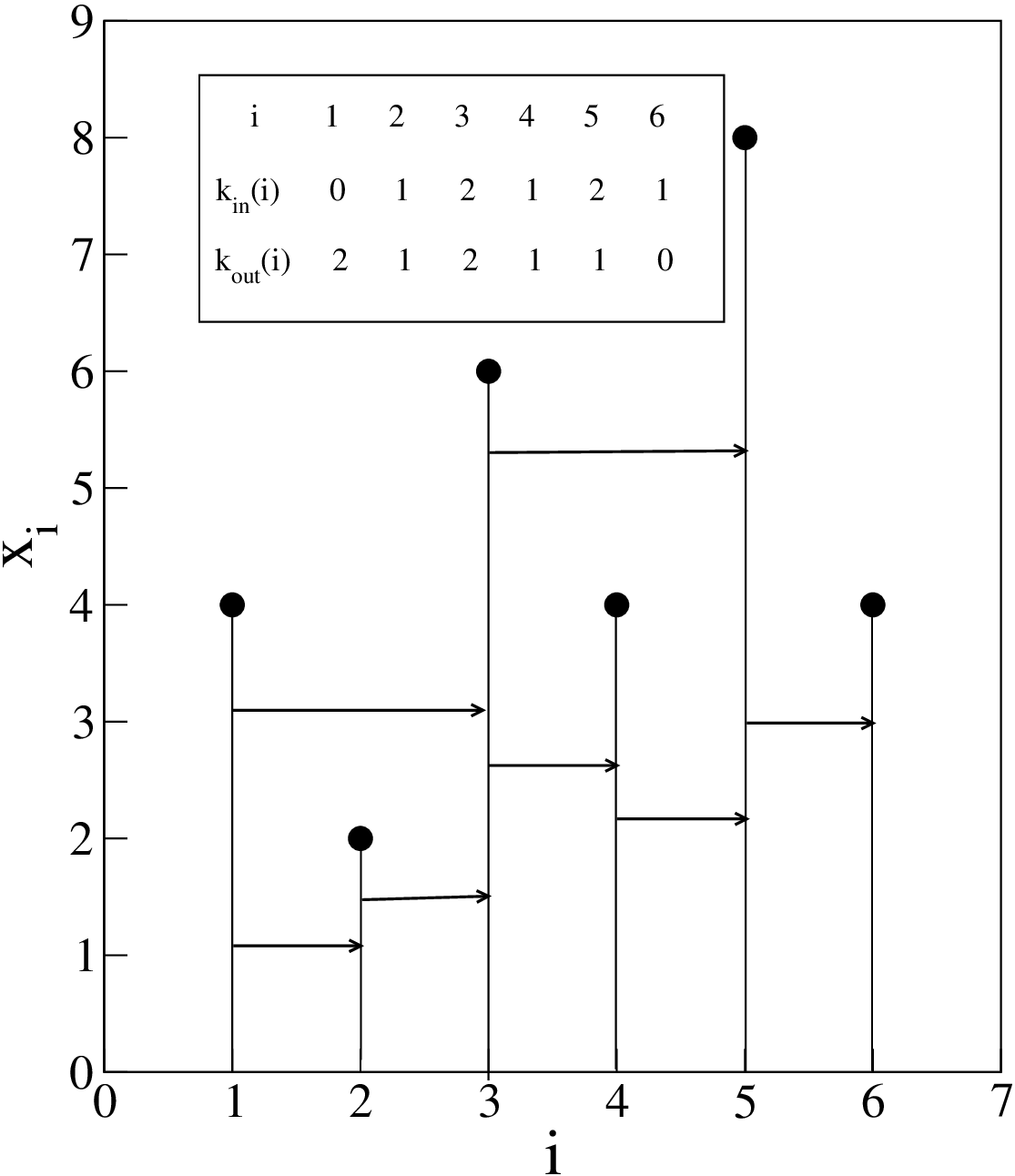}
\caption{ Graphical representation of a time series of $6$ data points.
Arrows show the `visibility'. Ingoing degree $k_{in}(i)$ and
outgoing degree $k_{out}(i)$ are mentioned in the inset.}
\label{fig1}
\end{figure}

Let ${x_i, i=1,2,...,N}$, be a time series such that
$x_{i+1}=x_i + \delta t$, where $\delta t$ is the sampling time.
Fig. \ref{fig1} shows a time series of $6$ data points.
As per DHVg algorithm \cite{laca12},
each data point is treated as a node of a graph. Two nodes
$i$ and $j$ are said to be connected if
\begin{eqnarray}
x_i, x_j > x_n \forall n; i<n<j.
\end{eqnarray}
For example, in Fig. \ref{fig1}, nodes $1$ and $3$ are connected
whereas nodes $1$ and $5$ are not connected.
Now an ingoing degree for $i^{th}$ node, $k_{in}(i)$ is defined
as the number of links of that node with other past nodes.
Similarly an outgoing degree $k_{out}(i)$ is defined as the
number of links with future nodes.
Corresponding to each node  of time series
shown in the Fig. \ref{fig1}, $k_{in}$ and $k_{out}$
are written in the inset. Now degree $k(i)$ of $i^{th}$
node can be written as
\begin{eqnarray}
k(i)=k_{in}(i)+k_{out}(i).
\end{eqnarray}
The degree distribution of a graph gives the probability of an arbitrary
node to have degree $k$ \cite{newm03}.
Let $P_{in}(i)$ and $P_{out}(i)$ denote the probability
distributions of `in' and `out' degree distributions respectively.
\begin{figure}
\vspace{1.2cm}
\centering
\includegraphics[scale=.5]{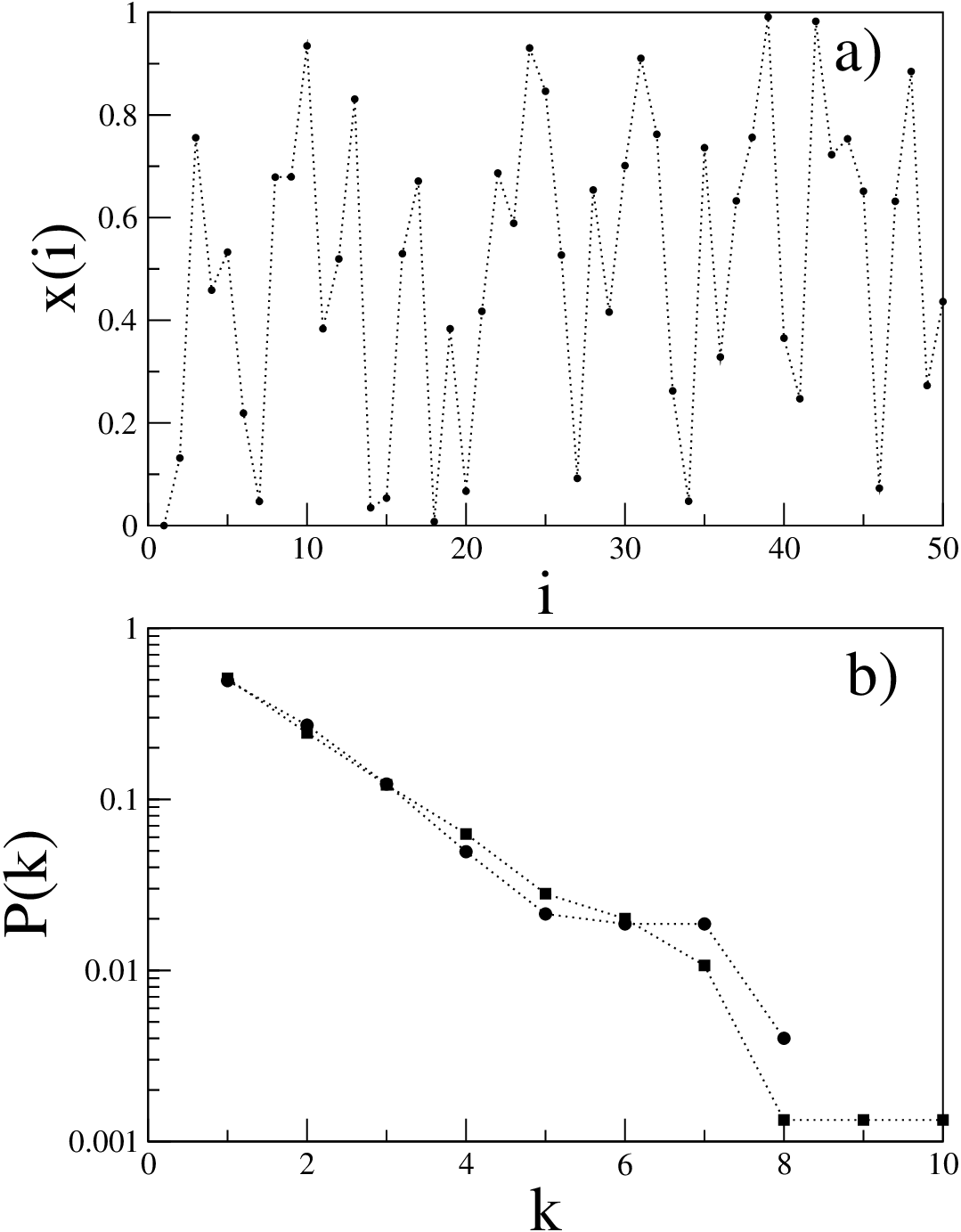}
\caption{(a)Random numbers ($50$ points) uniformly
distributed in the range $[0-1]$. (b)\textit{In} ($\CIRCLE$)
and \textit{out} ($\blacksquare$) degree distributions for
the time series shown in (a) (for $750$ data points).}
\label{fig2}
\end{figure}

\opencite{laca12} suggested that Kullback-Leibler divergence (KLD) 
\cite{cove06} between the \textit{in} and
\textit{out} degree distributions,
\begin{eqnarray}
D(P_{out}\parallel P_{in})=\sum_{i}{P_{out}(i) \mbox{log} \frac{P_{out}(i)}{P_{in}(i)}}
\label{d},
\end{eqnarray}
can be used as a measure of the irreversibility of two
real-valued  stochastic series.
$D$ vanishes if both the probability distributions are equal.
The more distinguishable are $P_{out}$ and $P_{in}$ with
respect each to other, the larger is the parameter $D$ and hence
more time irreversible the series is.
\section{Irreversibility of a time series}
\begin{figure}
\centering
\vspace{2cm}
\includegraphics[scale=.4]{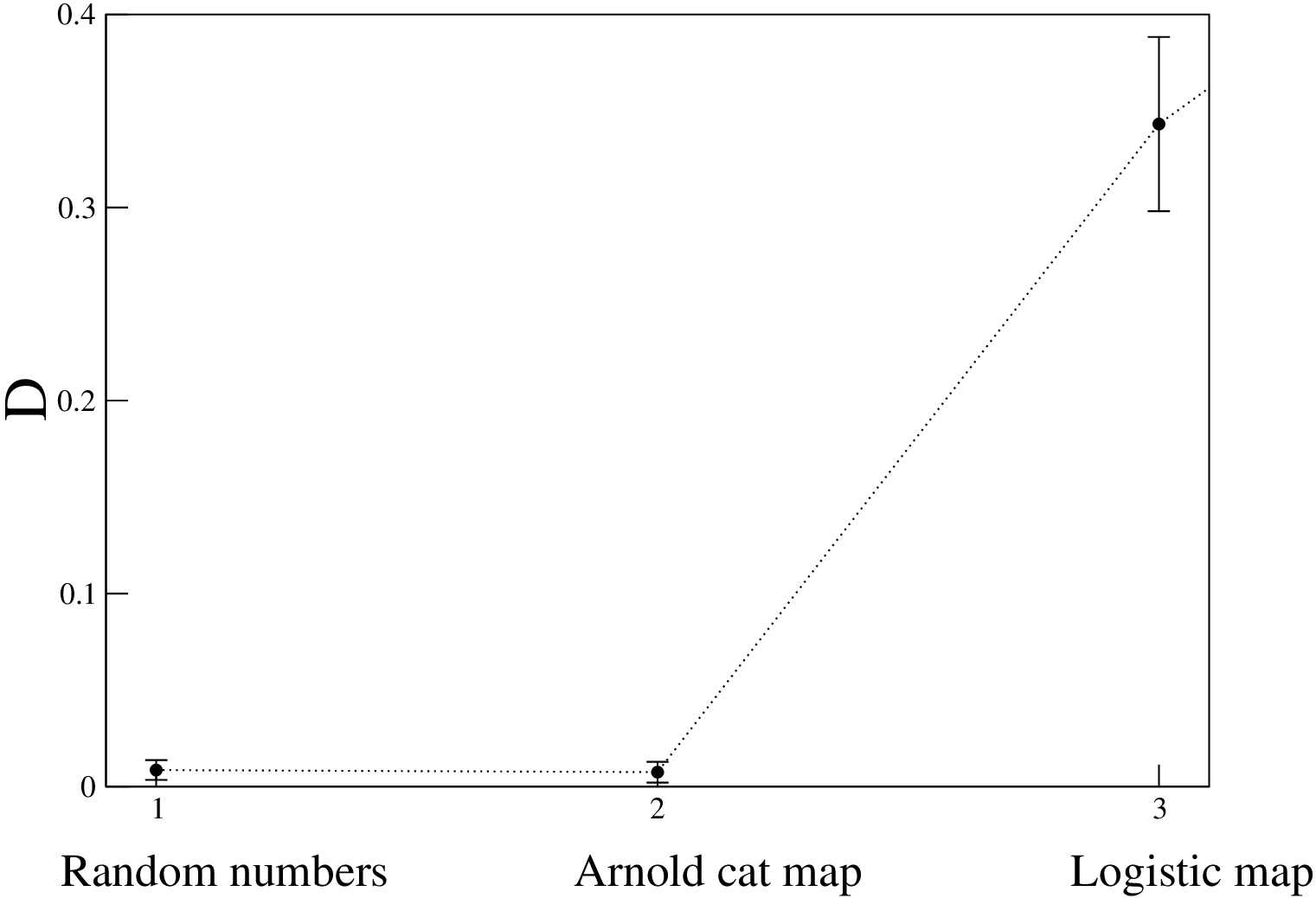}
\caption{Distance, $D$, for area preserving (e.g., uniformly distributed random numbers and
Arnold cat map) and dissipative (e.g., logistic map)) with 
 $750$ data points each.
1 $\sigma$ error bars are shown where $\sigma$ is the standard deviation.}
\label{fig3}
\end{figure}

In order to understand the time irreversibility in
a system using DHVg, we take few examples.
In all the examples we take time series of length $750$
data points as the shortest solar wind time series
available is  consists of approximately $750$ points.
First we generate a time series of uniformly distributed random
numbers in the range $[0,1]$. Fig. \ref{fig2} (a) shows first
$50$ points of this time series.
In Fig. \ref{fig2} (b) we show
the \textit{in} and \textit{out} degree distributions of
DHVg for this time series of $750$ data points.
We estimate $D$ for this data and find it to be very close to zero.
For second example, we take Arnold cat map \cite{arno68},
\begin{eqnarray}
\nonumber x_{t+1} = x_t + y_t~ \mbox {mod(1)},\\
y_{t+1} = x_t + 2 y_t~ \mbox {mod(1)}.
\end{eqnarray}
This is an example of a conservative chaotic system,
and the time series obtained from this map is
reversible in time. We  estimate the irreversibility parameter
$D$ for this time series and observed that it is  also very close to zero \cite{laca12}.
Now we consider the case of a dissipative and hence
time irreversible system, for example logistic map in the chaotic regime,
\begin{equation}
 x_{t+1} = 4 x_t (1-x_t).
\end{equation}
The estimated value of $D$ for the time series corresponding
to logistic map is significantly larger than that
the reversible ones.
We plot $D$ for all these time series, i.e.  obtained
from area preserving (e.g., random time series and
Arnold cat map) and dissipative (e.g., logistic map)
systems in  Fig. \ref{fig3}. This figure clearly
distinguishes the reversibility in these different type of systems.
These deterministic models are used just to demonstrate
the  usefulness of the DHVg analysis.

\begin{figure}
\vspace{1cm}
\centering
\includegraphics[scale=.4]{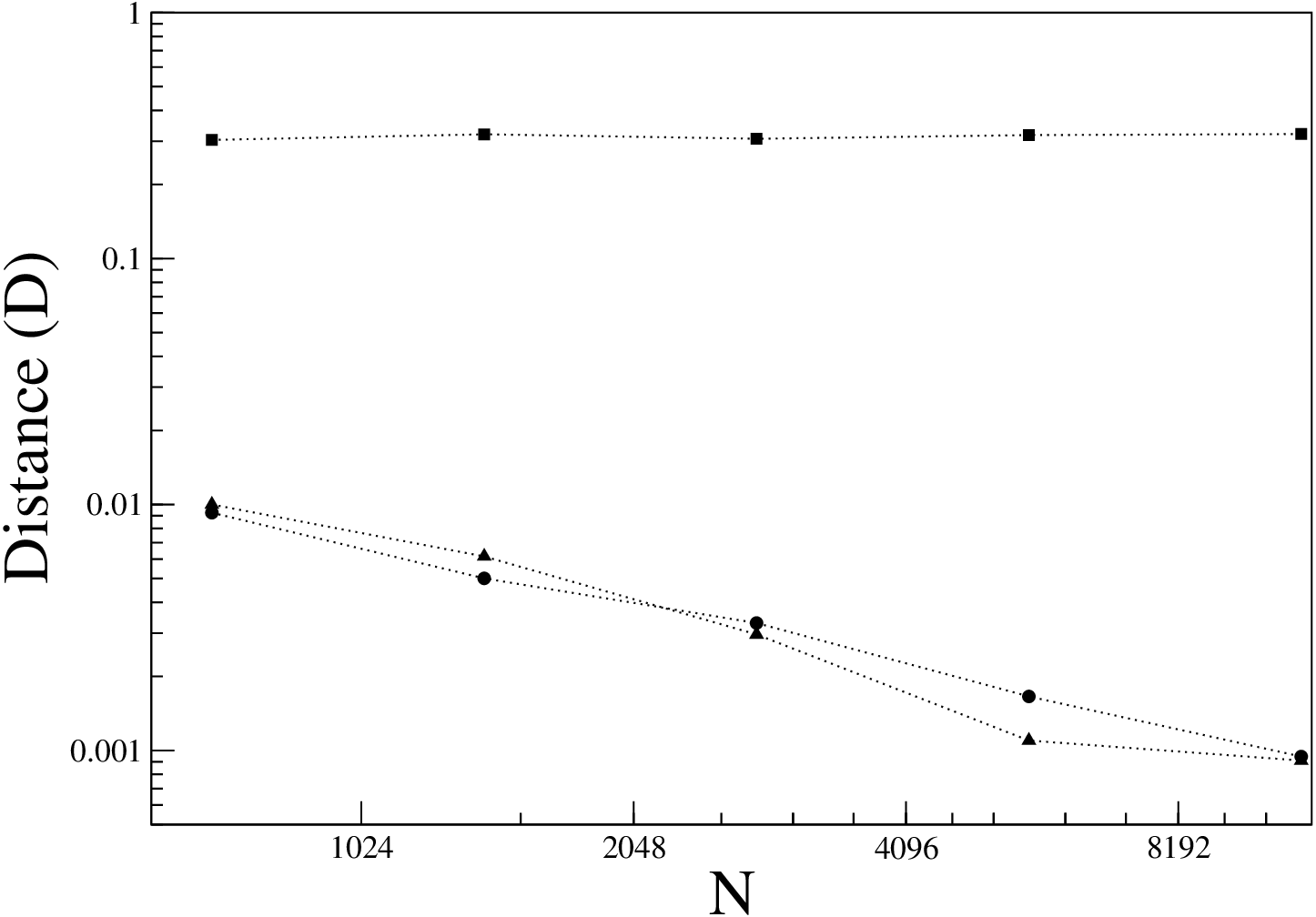}
\caption{Plot of $D[P_{out}\parallel P_{in}]$
of the graph associated to random number data ($\CIRCLE$),
data obtained from logistic map in the chaotic regime ($\blacksquare$),
and Arnold cat map ($\blacktriangle$) as a function of the series of length $N$.}
\label{fig4}
\end{figure}

In order to see  the effect of length of the time series  on
the irreversibility parameter we estimate $D$
for  different lengths.
Fig. \ref{fig4} shows the
plot of $D[P_{out}\parallel P_{in}]$ vs. lengths
of the time series obtained from different systems as considered above.
This plot shows that for a uniform random distributed time series ($\CIRCLE$)
and  time series generated from Arnold cat map ($\blacktriangle$),
the value of $D$ are small which
tends to zero as the length of the time series increases.
However for irreversible system (logistic map, $\blacksquare$) value of $D$ remains
significantly large as compared to that for
the reversible system (uniformly distributed random data or generated from Arnold map)
for different length of time series. This suggests that  $750$ data points
are thus sufficient to use DHVg measures to understand the irreversibility in time series.
Therefore we consider this number of data points to find the $D$ for 
solar wind data as discussed below.
\label{S-rev_irrev}
\section{Visibility graph analysis of solar wind data}
\begin{figure}
\centering
\vspace{2cm}
\includegraphics[scale=.4]{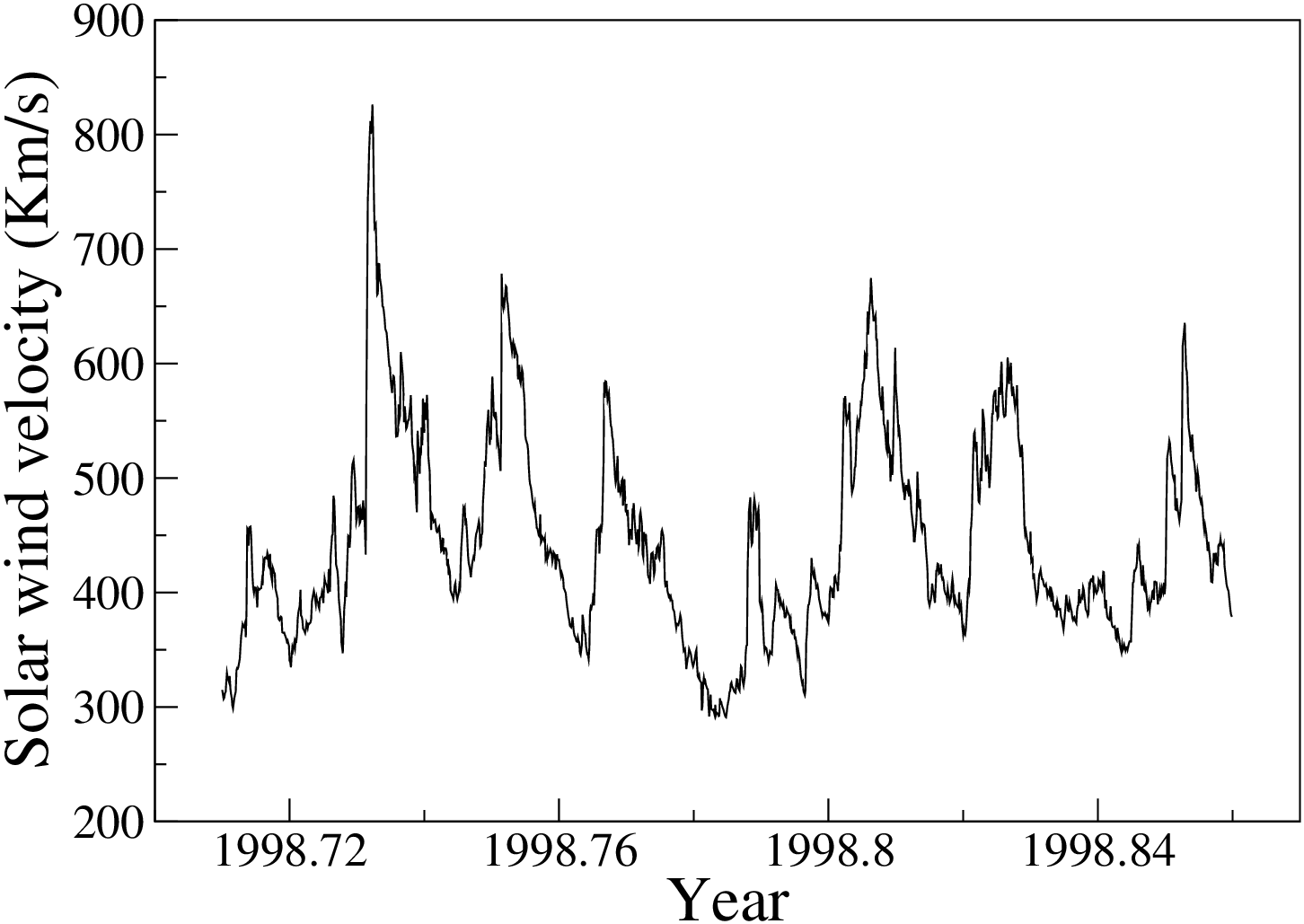}
\caption{Time series of solar wind velocity data obtained from ACE spacecraft
correspond to data set corresponds to S. No. $2$ of Table \ref{table_ace_1}.}
\label{fig5}
\end{figure}

\begin{figure}
\centering
\vspace{2cm}
\includegraphics[scale=.5]{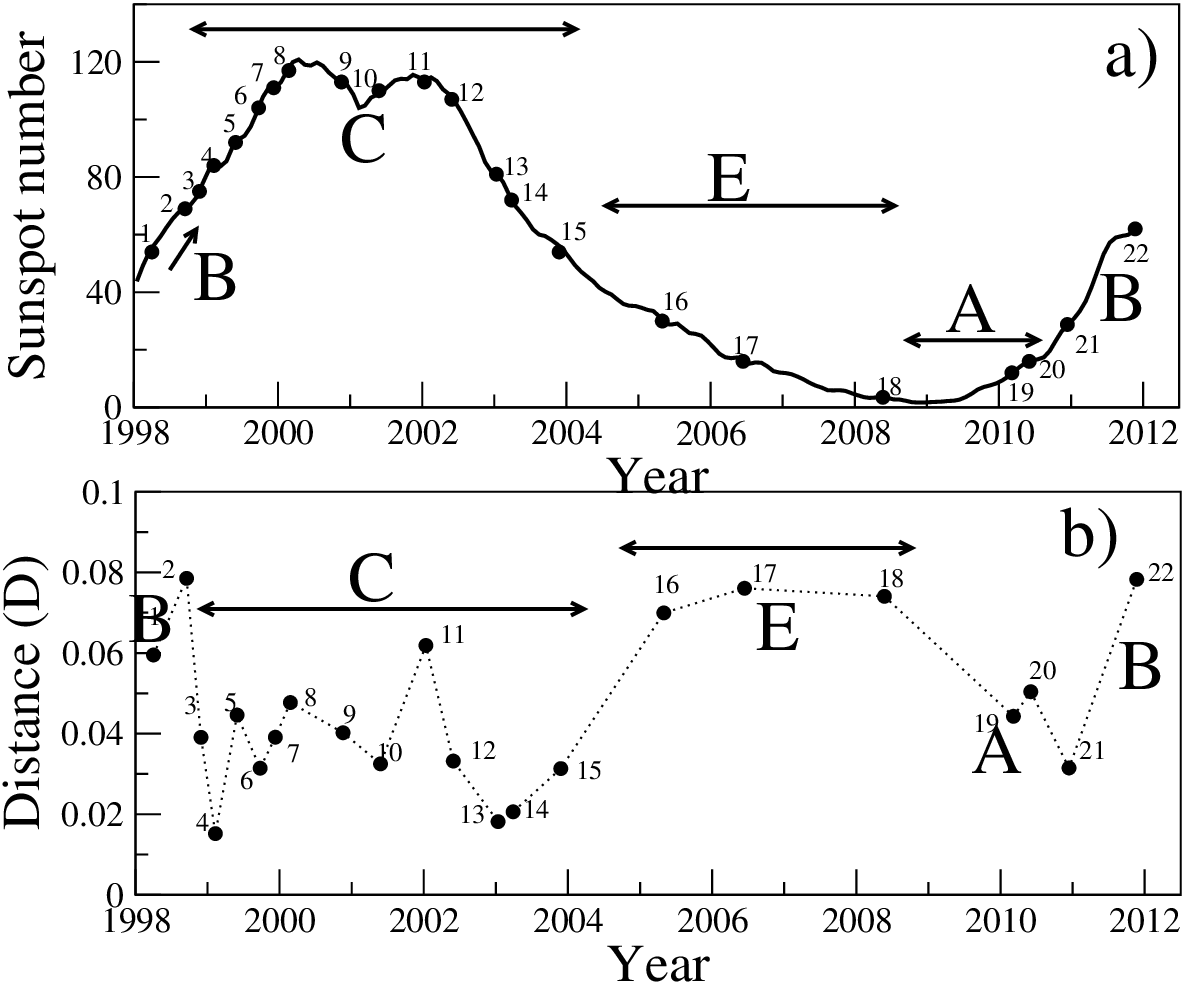}
\caption{ (a) Smoothed monthly averaged
sunspot number from $1996$ to $2012$.
Serial numbers $1-22$ correspond to
the  solar wind data sets given in Table \ref{table_ace_1}.
Arrows with symbols A, B, C, and E
indicate the low activity region,
ascending phase, high activity region,
and descending phase respectively.
(b) Distance, $D$ for solar wind velocity
corresponding to a) data.}
\label{fig6_a}
\end{figure}

The Advanced Composition Explorer was
launched on August, $1997$ to measure the elemental,
isotopic, and ionic charge-state composition
of nuclei from solar wind energies ($\sim 1$KeV nucl$^{-1}$)
to galactic cosmic-ray energies ($\sim 500$MeV nucl$^{-1}$) \cite{ston98}.
These data cover most of the solar activity
cycle $23$ and a part of solar cycle $24$.
We use hourly averaged solar wind velocity data at a distance of $1$ AU 
obtained from the Solar Wind Ion Composition Spectrometer
(SWICS) on ACE (http://www.srl.caltech.edu/ACE/ASC/level2/).
Due to data gap in the archive,
we split the data set into $22$ continuous small time series.
The details of the time series used are given in Table \ref{table_ace_1}.
Fig. \ref{fig5} (a) shows a typical time series of solar wind velocity
obtained from ACE spacecraft during $1998$.
In  Fig. \ref{fig6_a} (a) we plot smoothed monthly averaged sunspot number
(http://sidc.oma.be/sunspot-data/) during the same period.
Serial numbers show the corresponding solar wind data sets
on the smoothed sunspot index curve.
\begin{center}
\begin{table*}[h]
\caption{Initial time ($T$), of the $22$ time series of hourly
averaged solar wind velocity data measured
by the ACE spacecraft from $1998$ to $2012$.
Third column gives
time irreversibility parameter, $D$ (Eqn. \ref{d}).
The last column gives the smoothed monthly averaged sunspot numbers (SSN).}
\begin{tabular*}{0.75\textwidth}{|ccccc|}
\hline
S.No.&$T$&$N$&$D$&$SSN$\\
\hline
1&1998.25&1665&0.0595&55.0\\
2&1998.71&1314&0.0785&69.5\\
3&1998.91&789&0.0391&75.0\\
4&1999.11&1927&0.0152&84.6\\
5&1999.41&2716&0.0446&91.6\\
6&1999.73&1139&0.0314&102.3\\
7&1999.94&1404&0.0391&111.1\\
8&2000.15&3075&0.0477&116.8\\
9&2000.88&1580&0.0402&112.7\\
10&2001.40&1315&0.0324&109.4\\
11&2002.03&3329&0.0619&113.5\\
12&2002.41&3417&0.0332&107.5\\
13&2003.03&789&0.0182&80.8\\
14&2003.24&1840&0.0206&72.0\\
15&2003.90&2633&0.0313&55.7\\
16&2005.33&2541&0.0700&31.6\\
17&2006.45&2278&0.0761&16.3\\
18&2008.39&2108&0.0740&3.5\\
19&2010.18&1954&0.0443&12.3\\
20&2010.42&915&0.0504&15.9\\
21&2010.95&1182&0.0138&28.8\\
22&2011.89&867&0.0783&61.1\\
\hline
\end{tabular*}
\label{table_ace_1}
\end{table*}
\end{center}
\begin{center}
\begin{table*}[h]
\caption{Initial time ($T$), of the $15$ time series of $64$ sec
averaged solar wind velocity data measured
by the ACE spacecraft from $1998$ to $2012$.
Third column gives
time irreversibility parameter, $D$ (Eqn. \ref{d}).
The last column gives the smoothed monthly averaged sunspot numbers (SSN).}
\begin{tabular*}{0.75\textwidth}{|cccc|}
\hline
S.No.&$T$&$D$&$SSN$\\
\hline
   1&     1998.20&     0.0029&	53.5\\
   2&     1999.16&     0.0017&	84.6\\
   3&     2000.04&     0.0019&	112.9\\
   4&     2001.16&     0.0026&	104.8\\
   5&     2002.75&     0.0017&	94.6\\
   6&     2003.01&     0.0023&	80.8\\
   7&     2004.01&     0.0030&	52.0\\
   8&     2005.02&     0.0031&	34.6\\
   9&     2006.05&     0.0032&	20.8\\
  10&     2007.05&     0.0032&	11.9\\
  11&     2008.01&     0.0029&	4.2\\
  12&     2009.01&     0.0025&	1.8\\
  13&     2010.04&     0.0038&	9.3\\
  14&     2011.15&     0.0031&	33.4\\
  15&     2012.02&     0.0028&	65.5\\
\hline
\end{tabular*}
\label{table_ace_2}
\end{table*}
\end{center}
\begin{figure}
\includegraphics[scale=.5]{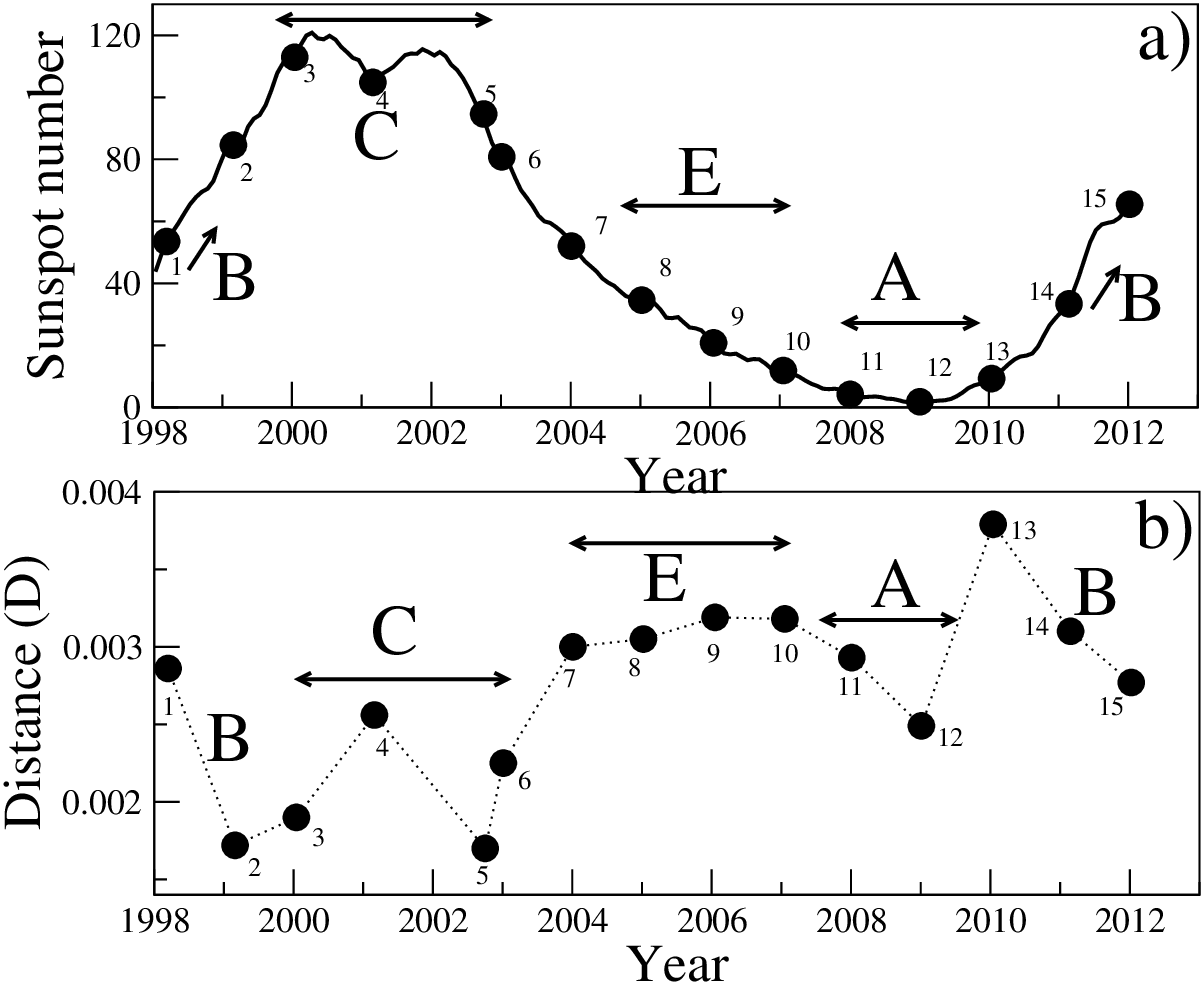}
\caption{ (a) Smoothed monthly averaged
sunspot number from $1996$ to $2012$.
Serial numbers $1-15$ correspond to
the  solar wind data sets given in Table \ref{table_ace_2}.
Arrows with symbols A, B, C, and E
indicate the low activity region,
ascending phase, high activity region,
and descending phase respectively.
(b) Distance, $D$ for solar wind velocity
corresponding to a) data.}
\label{fig6_b}
\end{figure}
We follow four phases of a solar activity cycle  namely,
low activity region, ascending phase, high activity region,
and descending phase.
These phases/regimes are
marked with alphabets, {\sl{A, B, C}} and {\sl{E}} respectively
in Figs. \ref{fig6_a},  \ref{fig6_b}, and \ref{fig8}.
Hourly averaged solar wind velocity data obtained from
ACE is such that time series $1$ and $2$ belong to
region B (ascending phase), time series $3$ to $15$
belong to region C (high activity region),
and time series $15$ to $18$ belong to
region E (the descending phase) of solar activity cycle $23$.
After time series $18$, a new activity cycle $24$
starts. Time series having serial numbers $21$ and $22$
resemble time series $1$ and $2$ of the
previous activity cycle respectively.
 
\begin{figure}
\centering
\includegraphics[scale=.5]{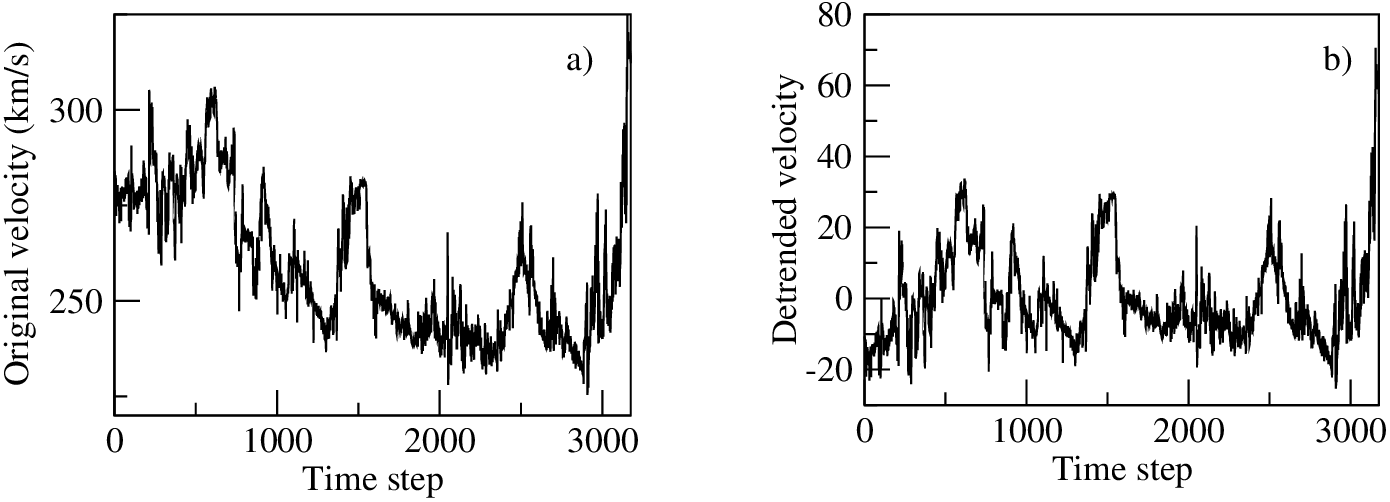}
\vspace{2cm}
\caption{(a) Time series of solar wind
velocity data obtained from Helios spacecraft
correspond to data set of S. No. $16$ of Table \ref{table_helios}.
(b) Detrended time series of a).}
\label{fig7}
\end{figure}
We estimate $D$ using DHVg analysis
for all these $22$ time series.
Since all the series are of different lengths,
we took $5$ overlapping windows of $750$ data points,
estimate $D$ for each window and then estimate the
average value of the $D$. Fig. \ref{fig6_a} (b)
shows variation of $D$ for solar wind velocity data
corresponding to the years $1998$ to $2012$.
We observe that estimated values of the
irreversibility parameter $D$ for
time series corresponding to region B,
\textit{i.e.}, ascending phase
of the activity cycle are larger than $D$ 
for region C, \textit{i.e.},
high activity region. $D$ fluctuates in a narrow range
for solar wind velocity data corresponding to this region.
(except time series $11$).
As the activity decreases (region E), estimated values
of $D$ increase sharply. They reach a peak
(for time series $17$ and $18$) at minimum activity.
On the onset of new activity cycle, $D$ decreases
and similar behavior  is observed
in the ascending phase of the next solar activity cycle although
the magnitude is different.

We also use $64$ sec averaged solar wind velocity data during
the same time period, \textit{i.e.}, from $1998$ to $2012$.
Table \ref{table_ace_2} gives the details of the time series used.
Fig. \ref{fig6_b} (a) shows the smoothed monthly averaged sunspot
data and Fig. \ref{fig6_b} (b) shows the variation of $D$
during the different phases of the solar activity cycle
for $64$ sec averaged data. A comparison of Fig. \ref{fig6_a} (b)
and Fig. \ref{fig6_b} (b) shows that the variation of $D$ shows similar
behavior for $64$ sec and hourly averaged data.

Since we do not have ACE data for high activity region
and descending phase of activity cycle $24$, we
consider another set of solar wind
data which is obtained from Helios space probes.
Helios space probes gathered solar wind data during part of
solar cycle $20$ and $21$ i.e., between $1975$ and $1983$.
They covered the heliocentric-distance from $0.28$ AU to $1.0$ AU.
We analyze radial velocity measured by the Helios spacecraft
at $0.3$ AU corresponding to the part of the solar activity cycle
$20$ and $21$. This data is obtained from \linebreak
$\mbox {http://sprg.ssl.berkeley.edu/impact/data\_browser\_helios.html}$.
The details of the time series used are
given in Table \ref{table_helios}.
Sampling time for each time series is $40.5$ sec.
In Fig. \ref{fig7} (a), we have shown a representative solar wind time
series obtained from Helios spacecraft.
The time series shows a decreasing
trend. Therefore, linear and quadratic
trends were subtracted
from the raw data \cite{mace97,gupt08}.
Fig. \ref{fig7} (b), shows the resulting detrended
time series for the data.
In  Fig. \ref{fig8} (a) we plot smoothed monthly
averaged sunspot number with time during the
years $1975-1983$. The position of
serial numbers corresponding to the solar wind data sets
is shown. We denote region A (time series from $1$-$7$),
region B (time series from $8$-$12$),
C (time series from $13$-$17$), and E (time series $17$
and $18$ on the activity cycle. 
In Fig. \ref{fig8} (b), we plot $D$ for all
the $18$ time series listed in Table \ref{table_helios}.
Fig. \ref{fig8} shows that the number of sunspots
does not change much for region corresponding
to time series $1$ to $9$, and $D$ fluctuates in
narrow range. During the ascending phase of the activity cycle
(time series $9$-$12$), $D$ increases and reaches a maximum.
In the high activity region (time series $13$-$17$)
$D$ remains almost constant.
During the descending phase of the activity cycle,
\textit{i.e.}, time series $17$ to $18$, $D$ increases sharply.
\begin{center}
\begin{table}[t]
\caption{Initial time ($T$), number of data points ($N$)
of the $18$ time series solar wind velocity data measured
by the Helios 2 spacecraft from $1976$ to $1982$. Third column gives
time irreversibility parameter, $D$ (Eqn. \ref{d}).
The last column gives daily averaged sunspot numbers (SSN).}
\begin{tabular*}{0.75\textwidth}{|ccccc|}
\hline
S.No.&$T$&$N$&$D$&$SSN$\\
\hline
1	&	1975.181	&	3515	&	      0.0064&20\\
2	&	1975.203	&	2938	&	      0.0123&22\\
3	&	1975.704	&	2979	&	      0.0078&17\\
4	&	1975.73 	&	3083	&	      0.0115&0\\
5	&	1976.205	&	2245	&	      0.0068&38\\
6	&	1976.287	&	1843	&	      0.0078&23\\
7	&	1976.791	&	1188	&	      0.0116&13\\
8	&	1977.205	&	2741	&	      0.0097&0\\
9	&	1977.287	&	3376	&	      0.0042&31\\
10	&	1977.791	&	1085	&	      0.0144&30\\
11	&	1978.286	&	2393	&	      0.0155&116\\
12	&	1978.371	&	1911	&	      0.0239&85\\
13	&	1979.370	&	1894	&	      0.0043&158\\
14	&	1980.372	&	2218	&	      0.0094&229\\
15	&	1980.454	&	2559	&	      0.0078&152\\
16	&	1981.456	&	3174	&	      0.0048&46\\
17	&	1981.458	&	2562	&	      0.0064&109\\
18	&	1982.453	&	1361	&	      0.0342&112\\
\hline
\end{tabular*}
\label{table_helios}
\end{table}
\end{center}
\begin{figure}
\vspace{2cm}
\centering
\includegraphics[scale=.5]{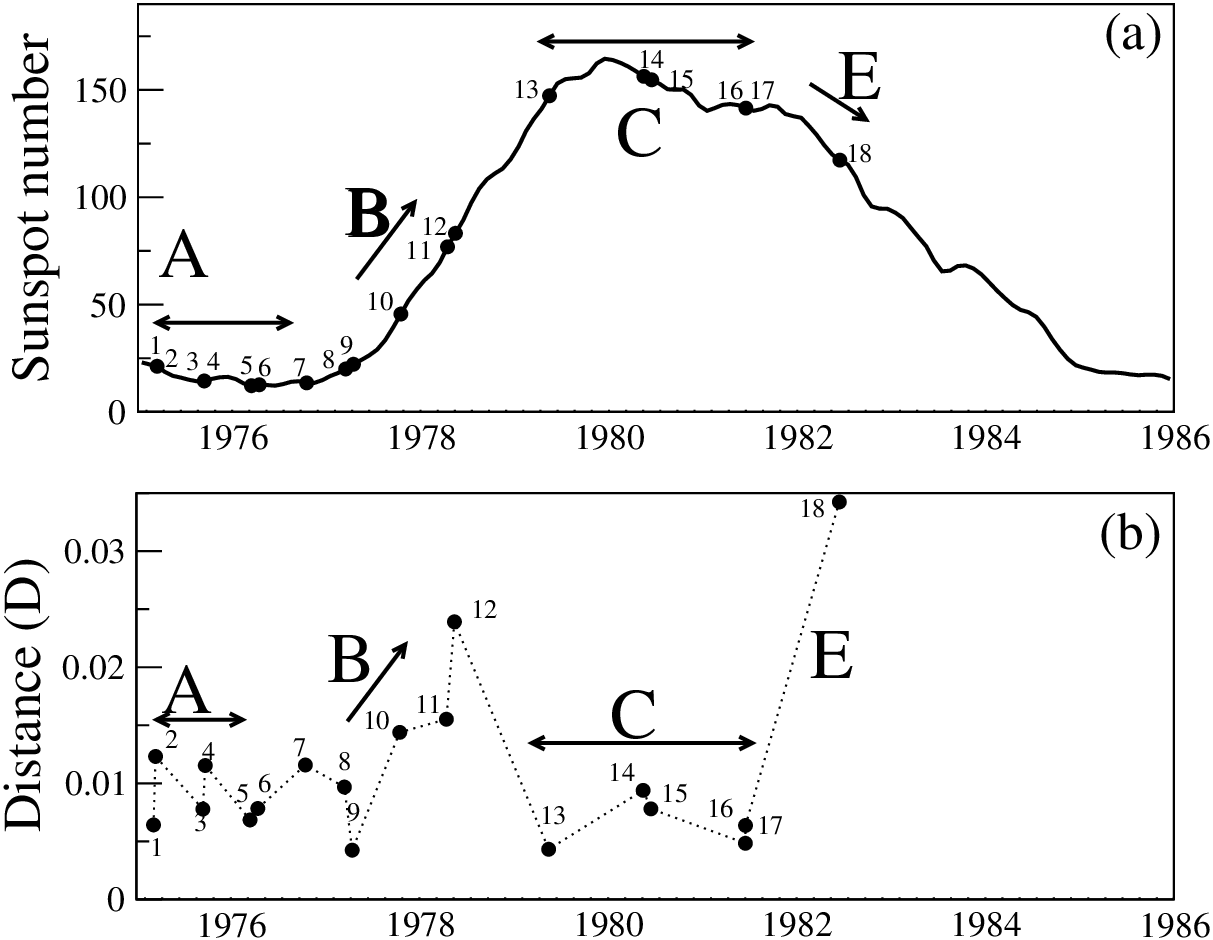}
\caption{ (a)  Smoothed monthly averaged
sunspot number from $1976$ to $1985$.
Serial numbers $1-18$ correspond to the  Helios solar
wind data sets given in Table \ref{table_helios}.
Arrows with symbols A, B, C, and E indicate the low activity region,
ascending phase, high activity region, and descending phase respectively.
(b) $D$ from DHVg analysis for solar wind
velocity corresponding to \ref{fig8} (a) data.}
\label{fig8}
\end{figure}
\label{S-solar_wind}
\section{Conclusions}
Internal properties, structure, and dynamics of the solar
wind vary spatially and temporarily \cite{balo08}.
Solar  wind is primarily divided into two components,
a slow wind with velocities and a fast
wind with velocities \cite{schw07}.
Fast solar wind originates from relatively dark and cool regions
in the corona, known as coronal holes \cite{feld76}.
Solar wind originates from above the more active regions on the Sun.
It is more variable as compared to fast solar wind \cite{krie73,woo97}.

Time series irreversibility gives the information about
the entropy production of the physical mechanism generating the
time series \cite{rold10}. In this paper, we present an analysis of solar
wind velocity data obtained from two spacecrafts,
ACE and Helios at distances $1$AU and $0.3$ AU
respectively. We obtained several different time series
of solar wind velocity from these two spacecrafts.
These time series belong to different phases of solar
activity cycles. To quantify the time irreversibility of
these time series, we use DHVg algorithm and estimate the
irreversibility parameter, $D$. A smaller value of $D$ indicates
a lesser time irreversible time series. Larger value of $D$
indicates that the system generating the time series is a
dissipative system.

We observe that $D$ increases in the ascending phase of an activity
cycle. After reaching to the maximum value, it  decreases
sharply in the high activity region of the activity cycle.
Further, in the descending phase, $D$ increases sharply,
saturates in the low activity region and repeats the similar
trend on the onset of new solar activity cycle.
This shows that solar wind velocity is more variable
during the high and low activity time,
while less variable during the ascending and descending
phases of the solar activity cycle.
Variation of $D$ over a solar activity cycle shows similar behavior
for both the distances, \textit{i.e.}, at $0.3$ AU and $1$ AU.
\label{S-conclusion}
\begin{acks}
The authors thank the ACE Science Center and instrument teams 
for making available the ACE data used here.
They thank Prof. Eckart Marsch, Max Planck Institute for Solar System Research,
Lindau, Germany, for help in obtaining the Helios data.
VS and AP thank CSIR for SRF, and Delhi University and DST
Govt. of India for financial supports respectively.
\end{acks}

\end{article} 
\end{document}